\documentclass[pre,
twocolumn,
aps,superscriptaddress,floats,showpacs]{revtex4}

\usepackage{graphicx}
\usepackage{dcolumn}
\usepackage{bm}
\usepackage{amssymb}
\usepackage{amsmath}
\usepackage{color}
\usepackage[T1]{fontenc}
\usepackage{yfonts}
\usepackage{color}

\begin{document}

\title{On the Schwinger limit attainability  with extreme power lasers}
\author{Stepan S. Bulanov}
\altaffiliation[Also at ]{Institute of Theoretical and Experimental Physics, Moscow 117218, Russia}
\altaffiliation[Present address: ]{University of California, Berkeley, 
CA 94720, USA}
\affiliation{University of Michigan, Centre of Ultrafast Optical Sciences, Ann Arbor, MI
48109, USA}
\author{Timur Zh. Esirkepov}
\affiliation{Kansai Photon Science Institute, JAEA, Kizugawa, Kyoto 619-0215, Japan}
\author{Alexander G. R. Thomas}
\affiliation{University of Michigan, Centre of Ultrafast Optical Sciences, Ann Arbor, MI
48109, USA}
\author{James K. Koga}
\affiliation{Kansai Photon Science Institute, JAEA, Kizugawa, Kyoto 619-0215, Japan}
\author{Sergei V. Bulanov}
\altaffiliation[Also at ]{Prokhorov Institute of General Physics, Russian Academy of Sciences, Moscow 119991, Russia}
\affiliation{Kansai Photon Science Institute, JAEA, Kizugawa, Kyoto 619-0215, Japan}

\begin{abstract}
High intensity colliding laser pulses can create abundant electron-positron pair plasma 
[A. R. Bell and J. G. Kirk, Phys. Rev. Lett. \textbf{101}, 200403 (2008)], 
which can scatter the incoming electromagnetic waves. This process can prevent reaching the critical field 
of Quantum Electrodynamics at which 
vacuum breakdown and polarization occur. Considering the pairs are seeded by the Schwinger mechanizm,
 it is shown that the effects of radiation friction 
and the electron-positron avalanche development
in vacuum depend on the electromagnetic wave polarization. For circularly polarized colliding pulses, 
which force the electrons to move in
circles, these effects dominate not only the particle motion but also the evolution of the pulses. While for linearly polarized
pulses, where the electrons (positrons) oscillate along the electric field, these effects are not as strong. 
There is an apparent analogy of
these cases with circular and linear electron accelerators with the corresponding constraining and reduced 
roles of synchrotron radiation losses.
\end{abstract}

\maketitle

The lasers nowadays provide one of the most powerful sources of electromagnetic (EM) radiation under laboratory conditions and thus
inspire the fast growing area of high field science aimed at the exploration of novel physical processes \cite{1}. Lasers have already
demonstrated the capability to generate light with the intensity of $2\times 10^{22}$W/cm$^2$ \cite{2} and projects to achieve
$10^{26}$W/cm$^2$ \cite{3} are under way. Further intensity growth towards and above $10^{23}$W/cm$^2$ will bring us to experimentally 
unexplored regimes. At such intensities the laser interaction with matter becomes strongly dissipative, due to
efficient EM energy transformation into high energy gamma rays \cite{1, 4}. These gamma-photons in the 
laser field may produce electron-positron pairs via the Breit-Wheeler process \cite{7}. Then the pairs accelerated by the laser
generate high energy gamma quanta and so on \cite{8}, and thus the conditions for the avalanche type discharge
are produced at the intensity $\approx$ 10$ ^{25}$ W/cm$^{2}$.  The occurrence of such "showers" was foreseen
by Heisenberg and Euler \cite{8a}. In Ref. \cite{9} a conclusion is made that depletion of the laser energy 
on the electron-positron-gamma-ray 
plasma (EPGP) creation could limit attainable EM wave intensity and could prevent approaching the critical quantum electrodynamics 
(QED) field. This field \cite{8a,10}
is also called the Schwinger field,  $ E_S=m_e^2c^3/e\hbar $ corresponding to the
intensity of $\approx$ 10$^{29}$W/cm$^2$. 

The particle-antiparticle pair creation by the Schwinger field cannot be described within the framework of perturbation theory and sheds
light on the nonlinear QED properties of the vacuum \cite{12}. Understanding the vacuum breakdown mechanisms is
challenging for other nonlinear quantum field theories \cite{13} and for astrophysics \cite{14}. Reaching this field limit has been
considered as one of the most intriguing scientific problems. Demonstration of the processes associated with the effects of
nonlinear QED, such as vacuum polarization and vacuum electron-positron pair production, will be one of the main challenges for extreme 
high power laser physics \cite{1, 17}.

In the present paper we discuss the attainability of the Schwinger field  with high power lasers. We compare the role of 
radiation dissipative effects in the motion of electrons (and positrons) produced via the Schwinger effect 
 and show their dependence on the EM wave polarization.

\begin{figure}[tbph]
\includegraphics[width=7.5cm,height=5cm]{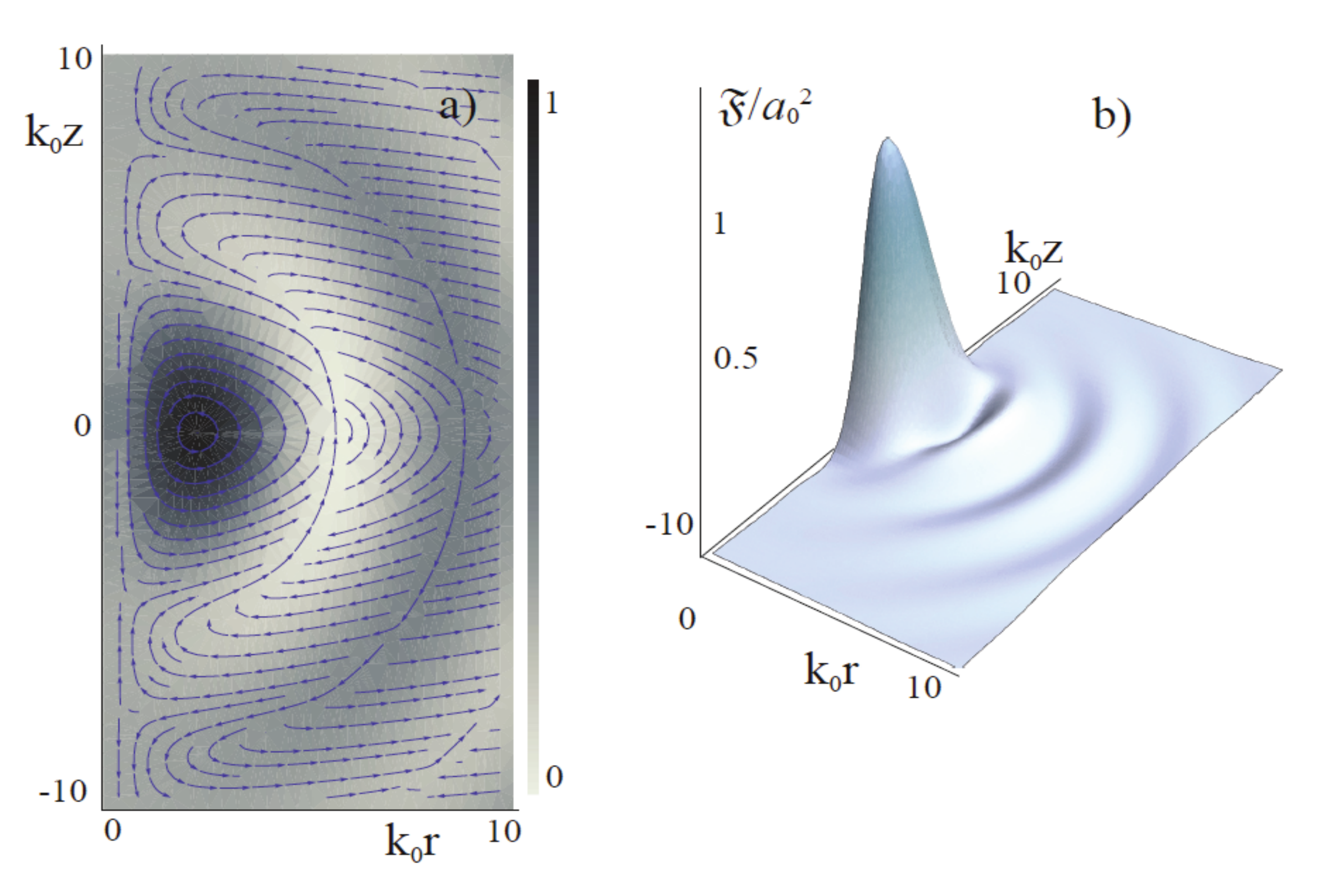}
\caption{ a) The vector field shows $r$- and $z$-components of the poloidal electric field in the $r,z$
 plane for the TM mode. The color density shows the toroidal magnetic field distribution, $B_{\phi}(r,z)$. 
 b) The first Poincare invariant $\mathfrak{F}(r,z)$.}
\label{FIG}
\end{figure}

Pair creation is determined by the Poincare invariants $\mathfrak{F}=({\bf E}^2-{\bf B}^2)/2$, 
$\mathfrak{G}=({\bf E}\cdot {\bf B})$
and requires the  first invariant $\mathfrak{F}$ be positive.
This condition can be fulfilled in the vicinity of the antinodes of colliding EM waves, or/and in the configuration 
formed by several focused EM pulses, \cite{20}. This EM configuration locally can be approximated by an oscillating 
TM mode with poloidal electric and toroidal 
magnetic fields. The magnetic field in spherical coordinates $R, \theta, \phi$ is given by 
\begin{equation}  \label{Bphi}
{\bf B}(R,\theta)={\bf e}_{\phi} \frac{a_0\sin(\omega_0 t)}{(8 \pi R)^{1/2}}  J_{n+1/2}(k_0 R) L^l_n(\cos\theta),
\end{equation} 
where $a_0=eE_0/m_ec\omega_0$, $k_0=\omega_0/c$,
$J_{\nu}(x)$ and $L^l_n(x)$ are the Bessel function and associated Legendre polnomials. 
The electric field is equal to ${\bf E}=i k_0 ( \nabla \times {\bf B})$.
 In cylindrical coordinates $r, \phi, z$ the $z$-component of the electric field oscillates in vertical direction, 
 $\sim a_0 \cos(\omega_0 t)$,
the $\phi$-component of the magnetic field vanishes on the axis being linearly proportional to the radius,  
$\sim (a_0/8) k_0 r \sin(\omega_0 t)$,
and the radial component of the electric field is relatively small, $\sim 0.1 a_0 k_0^2 r z \cos(\omega_0 t)$. 
The EM field and first Poincare invariant $\mathfrak{F}(r,z)$ are shown in Fig. \ref{FIG}.
We see that the EM field is localized in a region of width less than the laser wavelength, $\lambda_0=2 \pi /k_0$. 
The second invariant is equal to zero, $\mathfrak{G}=0$.

Using expression for the probability of electron-positron pair creation \cite{8a,10} and expanding $\mathfrak{F}(r,z)$ 
in the vicinity of its maximum
we find that the pairs are created in a small 4-volume near the electric field maximum with the characteristic 
size 
\begin{equation}  \label{deltaR}
\pi r_0^2 z_0 t_0 \approx  \frac{5^{3/2} \lambda_0^4}{16 \pi^5 c} \left(\frac{a_0}{ a_S}\right)^2.
\end{equation} 
Here, we introduce $a_S=eE_S/m_e \omega_0 c=m_e c^2/\hbar \omega_0$. 
Integrating over the 4-volume the probability of the pair creation \cite{25} we
 obtain the number of pairs produced per wave period, $(5^{3/2}/4 \pi^3) a_0^4 \exp(-\pi a_S/a_0)$,
 i. e. the first pairs can be observed for an one-micron wavelength laser intensity of the order of $2 \times 10^{27}$W/cm$^2$, 
which corresponds to  $a_0/a_S \approx 0.075$, i.e. a characteristic size, $r_0$, approximately equal to 0.04$\lambda_0$. 

In the region, where the magnetic field vanishes, the electron oscillates along the electric field. 
For an electron generated at small but finite radius $r_0 \ll \lambda_0$ the magnetic field bends its trajectory 
outwards. By solving the electron equations of motion linearized about the solution corresponding to ultrarelativistic electron 
oscillations in the $z$-direction, i.e. $a_0 \omega_0 t\gg 1$, we can find the electron trajectories, 
which are described in terms of modified Bessel functions. 
The instability growth rate is approximately equal to half the EM field frequency, $\omega_0/2$, 
i. e. the electron remains in the close vicinity of the zero-magnetic field region leaving it along the z-direction. 

The electron oscillating along the electric field  emits the high frequency EM radiation with the power 
$ \approx(2 \pi r_{e}/3\lambda_{0})\omega _e m_e
c^2\gamma_e^2$ proportional to the square of electron energy. In order to find the angular distribution and frequency spectrum of the
radiation in this case we should take into account its dependence on the retarded time: 
$t^{\prime }= t - {\bf n} \cdot {\bf r}(t) / c$. Here $ {\bf n} $ 
is the unit vector in the direction of observation and ${\bf r}(t)$ is the electron coordinate. Introducing the angle $\eta $
between vectors ${\bf n}$ and ${\bf r}(t)$, ${\bf n} \cdot {\bf r}(t) =  \vert {\bf r}(t)\vert \cos \eta $, we
can find that in the direction of electron oscillations, $\eta =0$, the radiation intensity vanishes. The maxima of the radiated power
correspond to the angle $\eta _m $, for large $\gamma _e $, inversely proportional to the particle energy: 
$\eta _m \approx 1 / 2\gamma _e $.

The Fourier components of the 4-vector potential of the EM field according to Ref. \cite{19} are
\begin{equation}  \label{eq6}
A^\mu (\omega ) = \frac{e}{R}\int\limits_{ - \infty }^{ + \infty }
{\frac{u^\mu }{c}\exp \left\{ {i\omega \left[ {t - \frac{1}{c}{\bf
n} \cdot {\bf r}(t)} \right]} \right\}dt} ,
\end{equation}
where $u^{\mu }=p^{\mu }/m_{e}\gamma _{e}$ is the four-velocity. ${\bf r}(t)={\bf e}_z(c/\omega _{0})\mathrm{Arcsin}
\left[ {\beta _{m}\sin (\omega _{0}t)}\right] $, and $\beta _{m}=a_{0} (1+a_{0}^{2})^{-1/2}$. Expanding the phase in expression 
(\ref{eq6}), 
$\Phi (t) = \omega \left\{ {t - \left( {\cos \eta / \omega _0 } \right) \mathrm{Arcsin} \left[ {\beta _m \sin (\omega _0 t)} \right]} \right\}$,  
over small parameters, $\gamma _{e,m}^{ - 1} $ and $\omega _0 t$, for $\eta =\eta _m \approx 1 / 2\gamma _{e,m} $, we obtain 
\begin{equation}  \label{PHI}
\Phi (t) \approx \omega \left[ {\left( {1 - \beta _m \cos \eta } \right)t 
+ \frac{\beta _m \cos \eta} {6 \omega _0 \gamma _{e,m}^2 } \left( {\omega _0 t} \right)^3} \right].
\end{equation} 
Using the Airy integral, we can find the $y$-component of the 4-vector potential of the EM field 
(\ref{eq6}) and the radiation power density. 
Since $\eta \sim 1/\gamma_{e,m} \ll 1$, and thus $\cos \eta - 1 \sim 1/ 2\gamma_{e.m}^2$,
the maximum frequency of the radiation emitted by the linearly oscillating electron is $\omega _m \approx 0.21\omega _0 \gamma_{e,m}^2 $.

To take into account the radiation friction we use equation of motion of a radiating electron \cite{19}.
We can estimate the regime where the radiation friction can become relatively large by comparing the energy losses with the maximal energy
gain of an electron accelerated by the electric field, $\dot {\mathcal{E}}^{( + )} \approx \omega _0 \,m_e c^2a_0 $,
i.e. $\omega _0 \,m_e c^2a_0=\varepsilon _{rad}\omega _0 m_e c^2\gamma_e^2$, where $\varepsilon _{rad}=4 \pi r_e/3 \lambda _0$,
with  $r_{e}=e^{2}/m_{e}c^{2}$.
As is apparent, although an electron moving along the oscillating electric field loses energy, radiation friction effects
may become important only at $a_0 = 2\varepsilon _{rad}^{ - 1} $, i.e. at the electric field $E_0 = 3m_e^2 c^4 / e^3$, which is of the
order of the critical electric field of classical electrodynamics (see also Ref. \cite{25}). This is 137 times larger than the field $E_S$.

In QED the charged particle interaction with EM fields is determined by relativistically and gauge invariant parameters
\cite{31} $\chi _{e}=[(F_{\mu \nu }p_{\nu })^{2}]^{1/2}/m_{e}cE_{S}$. The parameter, $\chi_{e}$, 
characterizes the probability of the gamma-photon 
emission by the electron with Lorentz factor $\gamma _{e}$. It is of the order of the ratio 
 $E/E_S$ in the electron rest frame of
reference. Another parameter, $\chi _{\gamma}=[(F_{\mu \nu }\hbar k_{\nu })^{2}]^{1/2}/m_{e}cE_{S}$, 
is similar to $\chi _{e}$ with the photon 4-momentum, 
$\hbar k_{\mu }$, instead of the electron 4-momentum, $p_{\mu }$. It characterizes the probability of the electron-positron pair
creation due to the collision between the high energy photon and EM field. QED effects come into play when the energy of a photon emitted by an
electron becomes comparable to the electron kinetic energy, i.e., for $\hbar \omega _m = m_e c^2\gamma _e $.
In a linearly polarized oscillating electric field the maximum frequency of emitted photons, $\omega _m $, is proportional $\gamma_0^2$, 
and, therefore, quantum effects should be incorporated into the theoretical description at the electron energy
corresponding to the gamma-factor $\gamma _Q^L = m_e c^2 / 0.21\,\hbar \omega _0 $, which is above the Schwinger limit.
We see that in the case of electron motion in a linearly polarized oscillating electric field neither radiation friction nor quantum
recoil effects are important.

Reaching the threshold of an avalanche type discharge with EPGP generation  discussed in Refs. \cite{8, 9} requires high enough
values of the parameters $\chi _e $ and $\chi _\gamma $ defined above because for $\chi _\gamma \ll 1$ the rate of the pair creation 
is exponentially small  \cite{33}, $W(\chi _\gamma ) \approx \alpha\left( m_e^2 c^4 / \hbar^2 \omega_{\gamma} \right)\chi _\gamma
\exp \left(  - 8 / 3\chi _\gamma  \right)$. In the limit $\chi _\gamma \gg 1$ the pair creation rate is given by $W(\chi
_\gamma ) \approx \alpha \left( m_e^2 c^4 / \hbar^2 \omega_{\gamma} \right)\left( \chi _\gamma \right)^{2 / 3}$ (for details
see Ref. \cite{31}). Here $\hbar \omega_{\gamma}$ is the energy of the photon which creates an electron-positron pair.

Since for $\gamma _e \ge \gamma _Q $ the photon is emitted by the electron (positron) in a narrow angle almost parallel to the
electron momentum with the energy of the order of the electron energy, the parameters $\chi _e $ and $\chi _\gamma $ are
approximately equal to each other. The parameter $\chi _e $ can be expressed via the electric and magnetic field as (see Ref. \cite{31})
\begin{equation}
\chi _{e}^{2}=
\left( 
\gamma_e \frac{{\bf E}}{E_S}+\frac{{\bf p} \times {\bf B} }{m_e c E_S}
\right)^2-
\left(
\frac{{\bf p} \cdot {\bf E} }{m_e c E_S}
\right)^2.  
\label{eq16}
\end{equation}

In order to find the threshold for the avalanche development we need to estimate the QED parameter $\chi_e$.
The condition for 
avalanche development corresponding to the parameter $\chi_e$ should become of the order of unity within one tenth of the EM field 
period (e.g. see Ref. \cite{9}). 
Due to the trajectory bending by the magnetic field the electron transverse momentum 
changes as $p_{\bot}\approx (a_0/16) k_0 r_0 (\omega_0 t)^2$,
where $k_0 r_0=(2.5 a_0/\pi a_s)^{1/2}$, Eq. (\ref{deltaR}). 
Assuming $\omega_0 t$ to be equal to 0.1 $\pi$, we obtain from Eq. (\ref{eq16}) 
that  $\chi_e$ becomes 
of the order of unity, i.e. the avalanche can start, at $a_0/a_S \approx 0.105$, which corresponds to the 
laser intensity $4 \times 10^{27}$W/cm$^2$. 
The radiation losses in this limit can be described as the synchrotron losses of an electron with the energy $\approx m_e c^2$
moving in the magnetic field $a_0 (k_0 r_0)/8$. Using formulae for synchrotron radiation \cite{19}, it is easy to show
that they do not become significant until $a_0 \approx 5\times 10^4$.  
At that limit the Schwinger mechanism provides approximately $5\times 10^5$ pairs per one-period. 

In the case of two colliding circularly polarized EM waves 
the resulting electric field rotates with frequency $\omega _{0}$ 
being constant in magnitude. The power emitted by
the electron is 
$\approx \varepsilon_{rad}\omega_{0}{{m}_{e}}c^{2}\gamma_{e}^{4}$. 
This is a factor of $\gamma _e^2 $ larger than in
the case of linear polarization.
The properties of radiation emitted by
 rotating electron  are well known from the
theory of synchrotron radiation \cite{19, 25} and from Ref.
\cite{26}. In the limit $\gamma_{e}\gg 1$ 
the emitted power is proportional to
the fourth power of the electron energy. The radiation is directed
almost along the electron momentum being localized
within the angle inversely proportional to the electron energy:
$\delta \eta \approx 1/\gamma _{e}$. The frequency spectrum given
by the well known expression \cite{19} has a maximum frequency,
$\omega _{m}=0.29\omega _{0}\gamma _{e}^{3}$, proportional to the
cube of the electron energy. This is a factor of $\gamma_{e}$ larger than in the
case of linear polarization.
For the electron rotating in the circularly
polarized colliding EM waves the emitted power becomes equal to the maximal
energy gain at the field amplitude $a_0 = a_{rad} = \varepsilon
_{rad}^{ - 1 / 3} $. For the laser wavelength $\lambda _0 =  0.8\,\mu$m  $\varepsilon _{rad} =
2.2\times 10^{ - 8}$. The normalized amplitude $a_{rad}$ is $\approx 400$ 
corresponding to the laser intensity $I_{rad} = 4.5\times 10^{23}$W
/ cm$^2$.

We represent the electric field and the electron momentum in the complex form: $E = E_y + iE_z
= E_0 \exp \left( {\ - i\omega _0 t} \right)$ and $p = p_y + ip_z = p_ \bot
\exp \left( {\ - i(\omega _0 t - \varphi )} \right)$, where $\varphi $ is
the phase equal to the angle between the electric field vector and the
electron momentum. In the stationary regime, when the electron
rotates with constant energy, the
equations for the electron energy, $\gamma _e = [1 + \left( {p_ \bot /
m_e c} \right)^2]^{1/2} $, and for the phase $\varphi $ have the form
\begin{equation}  \label{eq11}
a_0^2 = \left( {\gamma _e^2 - 1} \right)\left( {1 + \varepsilon _{rad}^2 \gamma _e^6 } \right)
 \,\, {\rm and }\,\,
\tan \varphi = - \frac{1 }{ \varepsilon _{rad} \gamma _e^3} .
\end{equation}
In the limit of weak radiation
damping, $a_0 \ll \varepsilon _{rad}^{ - 1 / 3} $, the absolute value of
the electron momentum is proportional to the electric field magnitude, $p_ \bot
= m_e ca_0 $, while in the regime of dominant radiation damping effects, i.e.
at $a_0 \gg \varepsilon _{rad}^{ - 1 / 3} $, it is given by $p_ \bot = m_e
c\left( {a_0 / \varepsilon _{rad} } \right)^{1 / 4}$.
For the momentum dependence given by this expression the power radiated by
an electron is $P_{\gamma ,C} = \omega _0 m_e c^2 a_0$,
i.e. the energy obtained from the driving electromagnetic wave is
completely re-radiated in the form of high energy gamma rays. At $a_0
\approx \varepsilon _{rad}^{ - 1 / 3} $ we have for the gamma photon energy
$\hbar \omega _\gamma = 0.29\hbar \omega _0 a_{rad}^3  \approx 0.45\hbar \omega _0 \left( {mc^3 / e^2} \right)$. 
For example, if $\lambda _0 \approx 0.8\,\mu$m and $a_0 \approx 400$ the circularly polarized laser pulse of
intensity $I_{rad} = 4.5\times 10^{23}\,W/cm^2$ generates a
burst of gamma photons of energy about 20 MeV with the duration
determined either by the laser pulse duration or by the decay time
of the laser pulse in a plasma.

Since in the case of circular polarization $\omega _m $ is proportional to the
cube of electron gamma-factor quantum effects should be incorporated
into the theoretical description at $\gamma _e \approx \gamma _Q^C = (m_e
c^2 / 0.29\,\hbar \omega _0 )^{1/2} \approx  1300$. For $\gamma _e = a_0 $ this limit is
reached at the intensity of $\approx 3.4\times 10^{24}$W/cm$^2$. The electron 
motion should be described within the framework
of quantum mechanics. These effects change the radiative loss
function (see Ref. \cite{31}). In the quantum regime, it is
necessary to take into account not only radiative damping effects but also
recoil momentum effects, which change the direction of motion of the
electron because the outgoing photon carries away the momentum $\hbar k_m =
\hbar \omega _m / c$. 

In the regime when the radiation friction
effects are important, i.e. when 
$a_0 \gg \varepsilon _{rad}^{ - 1 /3} $, the angle $\varphi $ between the electron momentum 
and the electric field is small being equal to
$\left( {\varepsilon _{rad} a_0^3 } \right)^{ - 1 / 4}$, i. e. the electron moves almost in the electric
field direction. The
electron momentum is given by $p_ \bot = m_e c\left( {a_0 / \varepsilon
_{rad} } \right)^{1 / 4}$. This yields an estimation
 $\chi _e \approx (a_0 / a_S^2 \varepsilon _{rad} )^{1/2} $. This 
becomes greater than unity for $a_0 > \varepsilon _{rad} a_S^2 \approx
5.5\times 10^3$, which corresponds to the laser intensity equal to
$ 6\times 10^{25}$W/cm$^2$. In Ref. \cite{9} an avalanche
threshold intensity several times lower has been found neglecting
the effects of the radiation friction force (see also \cite{34}). 
However, the radiation
friction time is of the order of $t_{rad} = 1 / \omega _0 \left(
{\varepsilon _{rad} a_0^3 } \right)^{1 / 2}$, which for $a_0 \approx
5.5\times 10^3$ is approximately one tenth of the laser period.
Hence the radiation friction effects do not prevent the EPGP cascade
development for circularly polarized colliding waves. Such a
prolific electron-positron pair and gamma ray creation \cite{8}
should result in the EPGP generation.

While creating and then accelerating the electron-positron pairs the
laser pulse generates an electric current and EM field. 
The electric field induced inside the EPGP cloud  with a size of the order of the
laser wavelength, $\lambda _0 $
can be estimated to be $E_{pol} = 2\pi e(n_ + + n_ - )\lambda _0 $.
Here $n_ + \approx n_ - $ are the electron and positron density,
respectively. Coherent scattering of the laser pulse away from
the focus region occurs when the polarization electric field becomes
equal to the laser electric field. This yields for the electron and
positron density $n_ + \approx n_ - = E / 4\pi e\lambda _0 $. 
The particle number per $\lambda^3_0$ volume is about $a_0 \lambda_0/r_e$.
This is a factor $a_0 $ smaller than required for the laser energy depletion.

In conclusion, the high enough laser intensity pulse  with arbitrary polarization plus high enough  density of seed electrons, 
e.g. generated in the laser interaction with solid targets  can provide 
necessary and sufficient conditions for the avalanche development, \cite{8}. Instead, in vacuum, when the seed 
electrons(positrons) are created via the Schwinger mechanism, we see a fundamental difference between 
the circularly and linearly polarized waves. In the case of the circularly polarized EM wave the electron 
radiation is strong and the threshold for the avalanche is low enough for avalanche
starting at the laser intensity well below the Schwinger limit. Since, as noted in Ref. \cite{8}, 
the electron-positron avalanche parameters are insensitive to the seed electrons (positrons), 
the parameters of the Schwinger created pairs become hidden and can hardly be revealed.
Contrary to this, in the linearly polarized EM wave is more favorable for the realization and reaching of "pure" 
Schwinger electron-positron pair creation.
An electron moving along the electric field with velocity and
acceleration parallel to the field emits much fewer photons with 
substantially lower energy
neither experiencing the radiation friction nor quantum recoil
effects. We see an analogy of these cases with
circular and linear electron accelerators with the corresponding
constraining and reduced roles of synchrotron radiation losses. The
electron-positron pair creation in the Breit-Wheeler type process is
also suppressed because the key parameters $\chi _e $ and $\chi
_\gamma $ dependence on the electron and photon momentum, in the laser field 
with the same intensity,is much weaker. 

We thank S. G. Bochkarev, V. Yu. Bychenkov, P. Chen, E. Esarey, A. M. Fedotov, V. F.
Frolov, D. Habs, M. Kando, K. Kondo, G. Korn, N. B. Narozhny, W. Rozmus, H. Ruhl, and A. I. 
Zelnikov for  discussions.
We acknowledge support of this work from 
MEXT 
of Japan, Grant-in-Aid for Scientific
Research, No. 20244065 and
from NSF through the Frontiers in Optical and Coherent
Ultrafast Science Center at the University of Michigan.


\begin{thebibliography}{99}
\bibitem{34}  We note here 
that using the Landau-Lifshitz approximation for the radiation friction force 
in the limit $a_0>a_{rad}$ leads to the losses scaling as 
$\sim \varepsilon_{rad} a_0^2 \gamma_e^2$ instead $\sim \varepsilon_{rad} \gamma_e^4$ 
and to underestimation of the laser intensity at which the parameter $\chi_e$
becomes equal to unity.
\end{thebibliography}
\end{document}